\documentclass[11pt]{article}
\usepackage{fullpage}
\usepackage{cite}
\usepackage{amsmath}
\usepackage{amssymb}
\title{}
\date{}
\def\para{\\ [-2mm]}
\def\be{\begin{equation}}
\def\ee{\end{equation}}
\def\ba{\begin{eqnarray}}
\def\ea{\end{eqnarray}}
\def\nl{\nonumber\\}
\def\eqn#1{eq.~(\ref{#1})} \def\Eqn#1{Equation~(\ref{#1})}
\def\eqns#1#2{eqs.~(\ref{#1}) and~(\ref{#2})}

\def\ie{{i.e.}}
\def\viz{{viz.}}
\def\eg{{e.g.}}

\def\sym{ {\rm sym} }
\def\SmodZ{  {S_n /\mathbb{Z} } }
\def\SL2C{\mathrm{SL}(2,\mathbb{C})}
\def\SmodG{S_n /\GT}
\def\id{  {\mathsf{1}\kern -3pt \mathsf{l} } }
\def\range{ 1,2, \cdots, n }
\def\cAYM{ {\cal A}}
\def\AYM{ A}
\def\cAgv{ {\cal M}}
\def\cAdc{ {\cal A}^{{\rm scalar}} }

\def\cT{ c^{(T)} }
\def\nT{ n^{(T)} }
\def\dT{ d^{(T)} }
\def\MT{ M^{(T)} }
\def\GT{ G^{(T)} }
\def\tn { \tilde{n} }
\def\tc { \tilde{c} }
\def \Tr {\mathop{\rm Tr}\nolimits}

\begin{document}

\titlepage
\begin{flushright}
BOW-PH-158\\
\end{flushright}

\vspace{3mm}

\begin{center}
{\Large\bf\sf
Scattering equations  \\
and virtuous kinematic numerators\\ [1.5mm]
and dual-trace functions}

\vskip 1.5cm

{\sc
Stephen G. Naculich\footnote{
Research supported in part by the National Science
Foundation under Grant No.~PHY10-67961.}
}

\vskip 0.5cm
{\it
Department of Physics\\
Bowdoin College\\
Brunswick, ME 04011, USA
}

\vspace{5mm}
{\tt
naculich@bowdoin.edu
}
\end{center}

\vskip 1.5cm

\begin{abstract}
Inspired by recent developments on scattering equations, we present
a constructive procedure for computing symmetric, amplitude-encoded,
BCJ numerators for $n$-point gauge-theory amplitudes, thus satisfying
the three virtues identified by Broedel and Carrasco.  We also develop
a constructive procedure for computing symmetric, amplitude-encoded
dual-trace functions $\tau$ for $n$-point amplitudes.  These can be
used to obtain symmetric kinematic numerators that automatically
satisfy color-kinematic duality.  The $S_n$ symmetry of $n$-point
gravity amplitudes formed from these symmetric dual-trace functions
is completely manifest.  Explicit expressions for four- and five-point
amplitudes  are presented.

\end{abstract}

\vspace*{0.5cm}

\vfil\break
\section{Introduction}
\setcounter{equation}{0}

Recent exciting work by Cachazo, He, and 
Yuan \cite{Cachazo:2013iaa,Cachazo:2013gna,Cachazo:2013hca,Cachazo:2013iea} 
has shed new light on the color-kinematic duality of gauge-theory amplitudes
and the double-copy construction that relates gauge-theory and gravity amplitudes \cite{Bern:2008qj,Bern:2010ue}.
Bern, Carrasco, and Johansson showed that tree-level gauge-theory 
and gravity amplitudes may be expressed as sums over cubic diagrams
\be
\cAYM ~=~ \sum_i {c_i ~ n_i \over d_i }, \qquad\qquad
\cAgv ~=~ \sum_i {\tn_i ~ n_i \over d_i }
\label{amps}
\ee
where the kinematic numerators $n_i$ are chosen to possess the same symmetries as
the color factors $c_i$.
Cachazo et al. \cite{Cachazo:2013iea}
extended the gauge-gravity nexus to include 
amplitudes of the theory of massless scalars in the adjoint representation
of $U(N) \times U(\tilde{N})$ with cubic interactions 
\be
\cAdc ~=~ \sum_i {\tc_i ~ c_i \over d_i }
\ee
and they showed that tree-level $n$-point scattering amplitudes for all three theories may be 
expressed in terms of an integral over the space of $n$ marked points on a sphere, $\{ \sigma_j ~|~ j=1, \cdots n\}$.
For example, the double-color scalar theory amplitudes may be expressed as\footnote{The definition of $m(\alpha|\beta)$ 
here differs from that of ref.~\cite{Cachazo:2013iea} by an overall sign when $n$ is even.}
\ba
\cAdc 
&=&
\sum_{\alpha, \beta} 
\Tr[\alpha]
~m(\alpha|\beta) ~
\widetilde{\Tr}[\beta]
\\
\Tr [\alpha] 
&=&
{\rm Tr}({T}^{\textsf{a}_{\alpha(1)}}{T}^{\textsf{a}_{\alpha(2)}}\cdots {T}^{\textsf{a}_{\alpha(n)}}) 
\\
m(\alpha|\beta)\ &=& (-1)^{n-1} ~\int \frac{d\,^n\sigma}{\textrm{vol}\,\SL2C}
\frac{\prod'_j 
~ \delta \left(\sum_{k\neq j} \frac{s_{jk}}{\sigma_{j,k}} \right)
}{(\sigma_{\alpha(1),\alpha(2)}\cdots\sigma_{\alpha(n),\alpha(1)})
(\sigma_{\beta(1),\beta(2)}\cdots\sigma_{\beta(n),\beta(1)})}
\label{local}
\ea
where $\sigma_{j,k}= \sigma_j - \sigma_k$
and $s_{jk}=( p_j + p_k)^2$ with $p_j$ the momenta of the external particles. 
The delta function localizes the integral (\ref{local}) on the solutions of the scattering 
equations \cite{Cachazo:2013iaa,Cachazo:2013gna}
\be
\sum_{k\neq j} \frac{s_{jk}}{\sigma_{j,k}} ~=~ 0 \quad {\rm for} \quad j=1, \cdots n \,.
\label{scatt}
\ee
As these equations have, up to $\SL2C$ transformations,
$(n-3)!$ solutions $\{\sigma_j^{(I)}\}$,
the double-partial amplitude $m(\alpha|\beta)$ may 
expressed as a sum over solutions 
\be
m(\alpha|\beta)=
~(-1)^{n-1}~\sum_{I=1}^{(n-3)!} \frac{1}{(\sigma^{(I)}_{\alpha(1),\alpha(2)}\cdots\sigma^{(I)}_{\alpha(n),\alpha(1)})(\sigma^{(I)}_{\beta(1),\beta(2)}\cdots\sigma^{(I)}_{\beta(n),\beta(1)}){\det}'\Phi(\sigma^{(I)})}
\label{sumoversolutions}
\ee
demonstrating that $m(\alpha|\beta)$ has rank $(n-3)!$.
The amplitudes (\ref{amps}) for gauge theory and gravity 
may also be succinctly expressed in terms of sums over 
solutions\cite{Cachazo:2013hca,Cachazo:2013iea}.
Related work on the scattering equations includes 
refs.~\cite{Litsey:2013jfa,Adamo:2013tca,Monteiro:2013rya,Mason:2013sva,Chiodaroli:2013upa,Dolan:2013isa,Adamo:2013tsa,Gomez:2013wza,Kalousios:2013eca,Stieberger:2014hba,Yuan:2014gva,Weinzierl:2014vwa,Dolan:2014ega,Bjerrum-Bohr:2014qwa,He:2014wua,Kol:2014yua,Kol:2014zca,Geyer:2014fka,Schwab:2014xua}.
\para

Because the color factors $c_i$ in \eqn{amps} satisfy Jacobi identities 
and hence are not linearly independent, the kinematic numerators $n_i$ 
for a given $n$-point gauge-theory amplitude are not uniquely determined. 
This generalized gauge freedom can be used to require that $n_i$ satisfy
the same Jacobi identities as the color factors \cite{Bern:2008qj}.
As a consequence all kinematic numerators $n_i$ may be expressed in terms of 
an independent set ${\bf n}_{1\gamma n}$ associated with half-ladder diagrams.
Defining color-ordered amplitudes $\AYM_\alpha$ as 
the coefficients of the 
gauge-theory amplitude expressed in the trace basis
\be
\cAYM  
= \sum_{\alpha} \Tr[ \alpha] ~ \AYM_{\alpha}
\label{trace}
\ee
it can then be shown that the  half-ladder numerators must satisfy 
\be
\AYM_\alpha   ~=~  \sum_{\gamma} m( \alpha| 1 \gamma n) ~{\bf n}_{1 \gamma n }  \,.
\ee
Because there are $(n-2)!$ independent half-ladder numerators 
${\bf n}_{1 \gamma n}$ and because rank $m = (n-3)!$,
this equation cannot be uniquely inverted, meaning that 
${\bf n}_{1 \gamma n }$ still possesses some residual generalized gauge freedom
even after color-kinematic duality is imposed.
Many different representations of the kinematic numerators are possible, 
\eg, see refs.~\cite{Kiermaier,BjerrumBohr:2010hn,Mafra:2011kj}.
\para

In an effort to define an economical and natural representation for the numerators, 
Broedel and Carrasco \cite{Broedel:2011pd} enumerated 
three virtues that kinematic numerators would ideally possess: 
(1) color-kinematic duality (numerators obey the same symmetries 
as the associated color factors),  
(2) amplitude-encoding (external-state dependence is expressed in terms of 
color-ordered amplitudes $A_\alpha$), and 
(3) symmetry (numerator functions corresponding to diagrams with the same topology 
but different labelings of the external legs 
are all related by permutations of their arguments).
They then proceeded to construct virtuous numerators for four- and five-point 
tree-level amplitudes, and for six-point MHV amplitudes in four dimensions,
by assuming a general ansatz for the numerators and imposing functional constraints.
Their approach becomes impracticable, however, for larger values of $n$,
and they voiced the hope that a constructive procedure for virtuous numerators
for arbitrary $n$-point amplitudes could be found.
\para

In sec.~3 of this paper, we present a procedure to produce $n$-point kinematic numerators
satisfying all three virtues of Broedel and Carrasco.
We begin with a specific set of nonsymmetric numerators from 
ref.~\cite{Cachazo:2013iea} obtained using the properties of $m(\alpha|\beta)$.
By applying arbitrary permutations to the external legs,
we derive other nonsymmetric sets of numerators.
Finally, we generate a symmetric numerator by summing over all such representations.\footnote{Recently,  
Fu, Du, and Feng \cite{Fu:2014pya}
presented a different but apparently equivalent algorithm by obtaining symmetric numerators
from a KLT expression for the gauge-theory amplitude.}
Our expression manifestly satisfies all Jacobi identities and diagram symmetries 
for the numerators without having to invoke the BCJ relations for the color-ordered amplitudes.
We give explicit results for four- and five-point amplitudes, but the procedure is 
completely general.
\para

The double-copy construction relates gauge-theory to gravity amplitudes by replacing 
the color factor $c_i$ in \eqn{amps} with a function $\tn_i$ of kinematic variables \cite{Bern:2008qj}.
Bern and Dennen \cite{Bern:2011ia} 
suggested that, in a similar way, the gravity amplitude could be
obtained from the gauge-theory amplitude (\ref{trace}) 
by replacing the color trace factor $\Tr[\alpha]$ 
with a dual-trace function $\tau_\alpha$ of the kinematic variables
\be
\cAgv
= \sum_{\alpha} \tau_\alpha ~ \AYM_{\alpha}  \,.
\ee
It can then be shown that 
\be
\AYM_\alpha   ~=~  \sum_{\beta} m( \alpha| \beta) \tau_\beta  \,.
\label{AYMmtau}
\ee
Once again, because there are $(n-1)!/2$ independent $\tau_\alpha$ and rank $m = (n-3)!$, this 
equation cannot be uniquely inverted;  many choices of dual-trace functions are possible. 
\para

In ref.~\cite{Bern:2011ia}, Bern and Dennen 
presented explicit expressions for $\tau_\alpha$ 
in terms of ${\bf n}_{1 \gamma n }$ 
for four-, five-, and six-point amplitudes,
and verified the existence of such expressions through nine points,
but did not present a general procedure for arbitrary $n$.
Further progress was achieved by Du et al. \cite{Du:2013sha,Fu:2013qna}.
\para

In sec.~4 of this paper, we present a procedure to generate 
$n$-point dual-trace functions $\tau_\alpha$ that possess 
the three virtues identified by Broedel and Carrasco:
(1) they automatically lead to kinematic numerators that obey color-kinematic duality,
(2) they are expressed in terms of $A_\alpha$, \ie, they are amplitude-encoded, and
(3) they are symmetric:  a single function $\tau$ suffices to determine the
full set via permutations of its arguments.
We first identify a specific set of nonsymmetric dual-trace functions
that satisfy \eqn{AYMmtau},
apply arbitrary permutations to the external legs to obtain other representations,
and then generate a symmetric dual-trace function by summing over all such representations.
We present explicit expressions for four- and five-point functions,
but again the procedure is completely general.
\para

The expressions for $\tau_\alpha$ obtained in ref.~\cite{Bern:2011ia} 
were observed to obey Kleiss-Kuijf relations,
and the authors argued that this was a necessary condition 
for symmetric dual-trace functions expressed in terms of kinematic numerators.
The results for $\tau_\alpha$ obtained in this paper do not satisfy Kleiss-Kuijf relations;
they are expressed directly in terms of color-ordered amplitudes,
and so are able to avoid this requirement.
Thus, even after the virtues of Broedel and Carrasco are imposed, 
there remains some freedom in the dual-trace function.
\para

This paper is structured as follows.
Section 2 reviews the tree-level amplitudes of 
gauge, gravity, and double-color scalar theories,
and the color-kinematic dualities that relate them.
Section 3 describes the criteria of Broedel and Carrasco
for virtuous numerators, outlines our procedure for constructing them,
and presents explicit results for four- and five-point amplitudes.
Section 4 reviews the properties of the dual-trace functions of 
Bern and Dennen, and then outlines our procedure for constructing them,
again presenting explicit results for four- and five-point amplitudes.
Section 5 discusses open questions,
and an appendix contains the proof of the procedure used
to produce symmetric kinematic numerators and dual-trace functions.

\section{Review of color-kinematic duality} 
\setcounter{equation}{0}

In this section, we review tree-level 
amplitudes of gauge, gravity, and double-color scalar theories,
and the color-kinematic dualities that relate these various 
amplitudes. \para

We begin with the tree-level $n$-gluon amplitude 
$\cAYM (\range)$, where the arguments denote 
the momentum $p_j$, polarization $\varepsilon_j$, and color $\textsf{a}_j$ 
of the external particles $j=1, \cdots, n$.
The color dependence can be specified in terms of a set of
color factors 
$c_i$ obtained by sewing together cubic 
vertices $f_{\textsf{abc}} \equiv \Tr(T^\textsf{a} [ T^\textsf{b}, T^\textsf{c}] )$,
where $T^\textsf{a}$ denote generators in the fundamental representation of the 
color group.
The $n$-point amplitude is then expressed as a sum over all cubic diagrams  
\cite{Bern:2008qj}
\be
\cAYM (\range) 
~=~ \sum_i 
{c_i ~ n_i (\range) 
\over d_i (\range)} \,.
\label{gluonamp}
\ee
The denominator $d_i (\range)$ 
associated with the color diagram $c_i$
is a product of the diagram propagators, 
and is a function of the external momenta $p_j$.
The kinematic numerator $n_i (\range)$ associated with $c_i$ 
is a function of both $p_j$ and $\varepsilon_j$.
(As usual, terms in the Feynman diagram expansion arising 
from quartic vertices can be parceled into terms involving only cubic vertices.) \para

The color factors $c_i$ can be expanded in a trace basis
\be
c_i  ~=~
\sum_{\alpha \in \SmodZ}  M_{i \alpha} \Tr[\alpha] , \qquad \qquad
\Tr [\alpha] ~\equiv~ 
{\rm Tr}({T}^{\textsf{a}_{\alpha(1)}}{T}^{\textsf{a}_{\alpha(2)}}\cdots {T}^{\textsf{a}_{\alpha(n)}})  \,.
\label{colortrace}
\ee
The trace basis is independent provided we restrict the sum to permutations 
of $n$ indices modulo cyclic permutations.
The gauge-theory amplitude (\ref{gluonamp}) can be decomposed in this basis 
\ba
\cAYM  (\range)
&=& \sum_{\alpha \in \SmodZ} \Tr[ \alpha] ~ \AYM_{\alpha} (\range)  ,
\label{tracedecomp}
\\
\AYM_{\alpha} (\range)
&=& \sum_i 
{M_{i \alpha}  ~ n_i (\range) 
\over d_i (\range)}
\label{colorordered}
\ea
where the arguments of the color-ordered amplitudes
$\AYM_{\alpha} (\range)$ denote $p_j$ and $\varepsilon_j$.
Since the trace basis is independent, the color-ordered amplitudes
are well-defined and gauge invariant.
Because of the invariance of the full amplitude $\cAYM (\range)$ under 
permutations of the arguments and the independence of the 
trace basis,\footnote{See discussion in the appendix.} 
the color-ordered amplitudes are all related to one another
via  
\be
\AYM_\alpha (\range) ~=~ \AYM ( \alpha(1), \cdots, \alpha(n))
\ee
where $\AYM  (i,j,\cdots) \equiv \AYM_{12\cdots n}(i, j, \cdots) $,
the coefficient of 
${\rm Tr}(T^{\textsf{a}_{1}}T^{\textsf{a}_{2}}\cdots T^{\textsf{a}_{n}})$.
The color-ordered amplitudes are thus ``symmetric'' functions, 
in the sense of Broedel and Carrasco\cite{Broedel:2011pd}. \para

The color factors $c_i$ in \eqn{gluonamp}
are not independent but satisfy various Jacobi identities
\be
c_i + c_j + c_k ~=~ 0 
\label{cjacobi}
\ee
which can be expressed 
\cite{Naculich:2011ep,Edison:2011ta,Edison:2012fn}
as $\sum_i \ell_i c_i  = 0$,
where $\ell_i$ are left null vectors of the matrix $M_{i\alpha}$: 
  $\sum_i  \ell_i M_{i\alpha}=0$.
The matrix $M_{i\alpha}$ has rank $(n-2)!$ so there are $(n-2)!$ independent
color factors.
The matrix $M_{i\alpha}$ also possesses a set of right null vectors 
$\sum_\alpha  M_{i\alpha} r_\alpha=0$.
By \eqn{colorordered}, these give 
rise\cite{Naculich:2011ep,Edison:2011ta,Edison:2012fn}
to a set of constraints
$\sum_\alpha  A_{\alpha} r_\alpha=0$
on the color-ordered amplitudes, \viz, 
the Kleiss-Kuijf relations \cite{Kleiss:1988ne}. \para

Because of the linear dependence of the color factors $c_i$, 
the kinematic numerators $n_i (\range)$ in \eqn{gluonamp}
are not uniquely determined but can undergo 
what are termed generalized gauge transformations \cite{Bern:2010yg} 
without altering the amplitudes.
The insight of Bern, Carrasco, and Johannson \cite{Bern:2008qj}
is that there exists a generalized gauge choice 
for which the numerators satisfy the same Jacobi identities as the color factors
\be
n_i (\range) ~+~ n_j (\range) ~+~ n_k (\range) ~=~ 0  \,.
\label{njacobi}
\ee
We will refer to a set of kinematic numerators satisfying \eqn{njacobi} as BCJ numerators.
Such a choice is not unique: there remain residual generalized 
gauge transformations that preserve the Jacobi identities (\ref{njacobi}).  
\para

A subset of the color factors $c_i$ are the half-ladder diagrams,
labeled by $ \alpha \in S_n$:
\be
{\bf c}_{\alpha} ~=~  \sum_{\textsf{b}_1,\ldots,\textsf{b}_{n{-}3}} 
f_{\textsf{a}_{\alpha(1)} \textsf{a}_{\alpha(2)} \textsf{b}_1}\cdots f_{\textsf{b}_{n{-}3} \textsf{a}_{\alpha(n{-}1)} \textsf{a}_{\alpha(n)}} 
  ~=~ \Tr( 
T^{\textsf{a}_{\alpha(1)}} 
[T^{\textsf{a}_{\alpha(2)}} ,
[ \cdots [
T^{\textsf{a}_{\alpha(n-1)}} ,
T^{\textsf{a}_{\alpha(n)}} ] \cdots ]] ) \,.
\label{c-basis}
\ee
An independent set of color factors \cite{DelDuca:1999ha,DelDuca:1999rs}
consists of those half ladders with $\alpha(1)=1$ and $\alpha(n)=n$:
\be
{\bf c}_{1 \gamma n } ~\equiv~
{\bf c}_{1 \gamma(2) \cdots \gamma(n-1) n } 
, \qquad\qquad \gamma \in S_{n-2} 
\ee
whose expansion in the trace basis begins
\be
{\bf c}_{1 \gamma n }  
~=~ \Tr[ 1 \gamma(2) \cdots \gamma(n-1) n]  
~+~ (-1)^n \Tr[ n \gamma(n-1) \cdots \gamma(1) 1]  ~+~ \cdots 
\label{halfladderexpand}
\ee
where $+ \cdots$ denotes traces without $1$ and $n$ adjacent.
That the set ${\bf c}_{1 \gamma n }$ is independent 
follows from \eqn{halfladderexpand} together with the independence of the trace basis 
(modulo cyclic permutations).
That the set is complete was shown in ref.~\cite{DelDuca:1999rs} 
where an arbitrary color diagram was reduced to a linear combination 
$c_i = \sum_i \lambda_{i,\gamma} {\bf c}_{1 \gamma n }$
using the Jacobi identities. 
By matching the coefficients of $\Tr[ 1 \gamma(2) \cdots \gamma(n-1) n] $
on both sides of this equation,
one then establishes that 
\be
c_i ~=~ \sum_{\gamma \in S_{n-2}}  M_{i, 1\gamma n} {\bf c}_{1 \gamma n } \,.
\label{cMc}
\ee
Then \eqns{gluonamp}{colorordered} imply \cite{DelDuca:1999ha,DelDuca:1999rs}  
\be
\cAYM (\range) ~=~ \sum_{\gamma \in S_{n-2}}  
{\bf c}_{1 \gamma n } A(1, \gamma(2), \cdots, \gamma(n-1), n) \,.
\label{cgammaA}
\ee

\subsection{Double-color scalar amplitudes}

In their seminal paper \cite{Bern:2008qj},
Bern, Carrasco, and Johansson showed that,
given a set of BCJ numerators $n_i(\range)$, 
one can obtain tree-level gravity amplitudes\footnote{
Up to an overall factor depending on coupling strengths
that we suppress throughout this paper}
by replacing $c_i$ in \eqn{gluonamp} with the kinematic 
numerators $\tn_i(\range)$ of a second gauge theory (the double-copy procedure)
\be
\cAgv (\range) 
~=~ \sum_i 
{ \tn_i (\range) ~ n_i (\range) 
\over d_i (\range)}
\label{gravamp}
\ee
where the $\tn_i(\range)$ need not themselves satisfy the Jacobi 
identities \cite{Bern:2010yg}.
\para

An alternative double-copy procedure replaces the $n_i(\range)$ in 
\eqn{gluonamp} with the color factors $\tc_i$ of a second color group
\be
\cAdc (\range) 
~=~ \sum_i 
{\tc_i ~ c_i \over d_i (\range)} \,.
\label{scalaramp}
\ee 
Such an expression corresponds \cite{Cachazo:2013iea} 
to the $n$-point amplitude of a theory of massless scalar 
particles $ \phi^{\textsf{a}\textsf{a'}}$
in the adjoint of the color group $U(N) \times U(\tilde{N})$ 
with cubic interactions of the form
\be
f_{\textsf{abc}}\tilde f_{\textsf{a'b'c'}}\phi^{\textsf{a}\textsf{a'}}\phi^{\textsf{b}\textsf{b'}}\phi^{\textsf{c}\textsf{c'}}
\ee
where 
$f_{\textsf{abc}}$ and $\tilde f_{\textsf{a'b'c'}}$
are the structure constants of $U(N)$ and $U(\tilde{N})$.
Using 
\be
\tc_i  ~=~
\sum_{\alpha \in \SmodZ}  M_{i \alpha} \widetilde{\Tr}[\alpha] , \qquad \qquad
\widetilde{\Tr} [\alpha] ~\equiv~ 
{\rm Tr}({\tilde T}^{\textsf{a}_{\alpha(1)}}{\tilde T}^{\textsf{a}_{\alpha(2)}}\cdots {\tilde T}^{\textsf{a}_{\alpha(n)}}) 
\ee
together with \eqn{colortrace}, the double-color amplitude (\ref{scalaramp}) can be written as
\be
\cAdc (\range)
~=~
\sum_{\alpha \in \SmodZ}  \sum_{\beta \in \SmodZ}  
\Tr[\alpha]
~m(\alpha|\beta) ~
\widetilde{\Tr}[\beta]
\label{TmT}
\ee
where\footnote{This definition of $m(\alpha|\beta)$ differs by a 
sign from that of ref.~\cite{Cachazo:2013iea} when $n$ is even.}
\be
m(\alpha| \beta)
~=~ \sum_i 
{M_{i\alpha} M_{i\beta} \over d_i(\range) }, 
\qquad \alpha, \beta \in S_n \,.
\label{doublepartial}
\ee
The coefficients $A_\alpha$ of $\Tr[\alpha]$ of the gauge-theory amplitude (\ref{tracedecomp})
are sometimes termed ``partial amplitudes.''
In ref.~\cite{Cachazo:2013iea}, 
Cachazo, He, and Yuan dubbed $m(\alpha|\beta)$,
the coefficients of 
$\Tr[\alpha] \widetilde{\Tr}[\beta]$ in \eqn{TmT},
``double-partial amplitudes.''
They showed that $m(\alpha|\beta)$ computes the sum of all trivalent scalar diagrams 
that can be regarded both as $\alpha$-color-ordered and $\beta$-color-ordered, 
where each diagram's contribution is given by the product of its propagators. 
The double-partial amplitudes satisfy (on both sides) 
Kleiss-Kuijf relations 
$ \sum_\alpha r_\alpha m(\alpha|\beta) = \sum_\alpha m(\alpha|\beta)  r_\beta = 0$
where $r_\alpha$ are right null vectors of the rank $(n-2)!$ matrix $M_{i\alpha}$.  
\para

Using \eqn{cMc}, the double-color amplitude (\ref{scalaramp}) can also be expressed as 
\be
\cAdc (\range)
~=~  \sum_{\gamma \in S_{n-2}}  \sum_{\delta \in S_{n-2}}  
{\bf \tc}_{1 \gamma n } ~m( 1 \gamma n| 1 \delta n) ~{\bf c}_{1 \delta n }  \,.
\label{cmc}
\ee
The double-partial amplitudes in this equation, 
$m( 1 \gamma n| 1 \delta n)$,
are essentially the entries in the $(n-2)! \times (n-2)!$ matrix
considered in ref.~\cite{Vaman:2010ez}.
There it was shown by consideration of low values of $n$ that,
as a result of momentum conservation,
the rank of this matrix is $(n-3)!$.
Cachazo et al. showed that the double-partial amplitudes could alternatively be expressed
as \eqn{sumoversolutions}, which shows explicitly that 
$m(\alpha|\beta)$ has rank $(n-3)!$. 
It consequently possesses $(n-3)!-(n-2)!$ additional null vectors,
dependent on kinematic invariants,
which implies that the double-partial amplitudes satisfy 
BCJ relations \cite{Bern:2008qj}  in addition to Kleiss-Kuijf relations.

\subsection{Gauge-theory amplitudes} 

Just as the Jacobi identities for the color factors (\ref{cjacobi})
imply that they can be expressed as a linear combination of
independent half-ladder color factors (\ref{cMc}), 
so the Jacobi identities (\ref{njacobi})
imply that the kinematic numerators can be written
in terms of an independent basis of numerators  
${\bf n}_{1 \gamma n } (\range) $ associated with ${\bf c}_{1 \gamma n } $
\be
n_i (\range) ~=~ \sum_{\gamma \in S_{n-2}}  M_{i, 1\gamma n} 
~{\bf n}_{1 \gamma n } (\range) \,.
\label{nMn}
\ee
Then, analogous to  \eqn{cmc} for double-color scalar amplitude,
we can write the gauge-theory amplitude (\ref{gluonamp}) 
as
\be
\cAYM (\range)
~=~  \sum_{\gamma \in S_{n-2}}  \sum_{\delta \in S_{n-2}}  
{\bf c}_{1 \gamma n } ~m( 1 \gamma n| 1 \delta n) ~{\bf n}_{1 \delta n }  \,.
\label{cmn}
\ee
We can also substitute \eqn{nMn} into \eqn{colorordered} 
to obtain the color-ordered amplitudes
\be
\AYM_\alpha   (\range) 
~=~  \sum_{\gamma \in S_{n-2}}  
m( \alpha| 1 \gamma n) 
~{\bf n}_{1 \gamma n } (\range) \,.
\label{Amn}
\ee
As noted above, 
the matrix $m( \alpha| 1 \gamma n) $ has rank $(n-3)!$
and has $(n-2)! - (n-3)!$ additional\footnote{In addition, 
that is, to the null vectors of $M_{i\alpha}$ 
which give rise to the Kleiss-Kuijf relations.}
null vectors (dependent on kinematic invariants)
which give rise \cite{Vaman:2010ez,Boels:2012sy,Litsey:2013jfa} 
to the BCJ relations \cite{Bern:2008qj} among the color-ordered amplitudes.
\para

The kernel of the matrix corresponds to 
$(n-2)! - (n-3)!$ degrees of freedom of
residual generalized gauge transformations of the BCJ numerators
(i.e. generalized gauge transformations  that preserve the Jacobi constraints (\ref{njacobi})).
Because of this residual gauge freedom, 
\eqn{Amn} cannot be inverted 
to obtain unique expressions for BCJ numerators 
in terms of color-ordered amplitudes.
To invert \eqn{Amn}, we must first make a choice of gauge. 
One possible gauge choice is to set 
\be 
{\bf n}_{1\gamma(2)\cdots\gamma(n-1)n} = 0, \qquad \qquad \gamma(n-1)\neq n-1
\label{zeronumerators}
\ee
which allows us to restrict the sum in \eqn{Amn} to $S_{n-3}$.
\para

Now consider the subset of color-ordered amplitudes
\be
\AYM (1, \beta(2), \cdots, \beta(n-2), n, n-1) 
~=~  \sum_{\gamma \in S_{n-3}}  
m( 1 \beta n, n-1 | 1 \gamma n-1,n) 
~ {\bf n}_{1\gamma(2)\cdots\gamma(n-2)n-1,n} (\range)   \,.
\label{mbetagamma}
\ee
It was shown in ref.~\cite{Cachazo:2013iea}
by using KLT orthogonality \cite{Cachazo:2013gna} that 
the  $(n-3)! \times (n-3)!$ submatrix
appearing in \eqn{mbetagamma} is invertible, 
with the inverse given by
the (negative of the) momentum kernel\footnote{Here $\gamma,\beta\in S_{n-3}$ are permutations acting on 
labels $2,3,\ldots,n{-}2$; $\theta(r,s)_\beta=1$ if the ordering of $r,s$ 
is the same in both sequences of labels, $\gamma(2),\ldots,\gamma(n{-}2)$ 
and $\beta(2),\ldots,\beta(n{-}2)$, and zero otherwise.} \cite{BjerrumBohr:2010ta,BjerrumBohr:2010zb,BjerrumBohr:2010yc,BjerrumBohr:2010hn}
\be  
S[\gamma|\beta]~=~
\prod^{n{-}2}_{i=2}\left[s_{1, \gamma(i)}+\sum^{i{-}1}_{j=2} \theta(\gamma(j), \gamma(i))_{\beta} s_{\gamma(j),\gamma(i)}\right].
\ee
Thus the nonzero members of the independent basis of numerators 
can be expressed as \cite{Cachazo:2013iea} 
\be
{\bf n}_{1\gamma(2)\cdots\gamma(n-2)n-1,n} (\range) 
~=~ -~\sum_{\beta\in S_{n-3}}S[\gamma|\beta] 
\AYM (1,\beta(2), \cdots, \beta(n-2) ,n,n-1) \,.
\label{BCJnumerator}
\ee
The full set of (BCJ) numerators,
including the half-ladders not included in the independent set,
can then be obtained from \eqns{zeronumerators}{BCJnumerator} via \eqn{nMn}.
\para

\subsection{Gravity amplitudes} 

By the double-copy construction, we can replace 
${\bf c}_{1 \gamma n }$ in \eqns{cgammaA}{cmn} with
${\bf  n}_{1 \gamma n } (\range)$ 
to obtain the  gravity amplitude
\ba
\cAgv (\range)
&=& \sum_{\gamma \in S_{n-2}}  
{\bf n}_{1 \gamma n } (\range)
\AYM (1, \gamma(2), \cdots, \gamma(n-1),n) 
\nl
&=&  \sum_{\gamma \in S_{n-2}}  \sum_{\delta \in S_{n-2}}  
{\bf n}_{1 \gamma n } (\range) ~m( 1 \gamma n| 1 \delta n) ~{\bf n}_{1 \delta n } (\range) \,.
\ea
The specific choice of gauge in \eqns{zeronumerators}{BCJnumerator} can then
be used to write
\ba
&& \cAgv (\range) 
\nl 
&& ~~~
=
 \sum_{\gamma \in S_{n-3}}  
   \sum_{\delta \in S_{n-3}}  
\AYM (1, \gamma(2), \cdots, \gamma(n-2),n-1,n)
S[\gamma|\delta] 
\AYM (1,\delta(2), \cdots, \delta(n-2) ,n,n-1)
\nl
\ea
which is one possible form of the field-theory limit of the KLT relation \cite{Kawai:1985xq,Bern:1998sv,BjerrumBohr:2010hn}.
\para

\section{Virtuous kinematic numerators}
\setcounter{equation}{0}

As described in the previous section, 
even after requiring the kinematic numerators to satisfy color-kinematic duality,
a certain amount of generalized gauge freedom remains.
To fix this residual gauge freedom in a natural, economical,
and possibly unique way, 
Broedel and Carrasco \cite{Broedel:2011pd} identified three
desirable features that a set of numerators should possess:

\begin{enumerate}
\item
color-kinematic duality:
BCJ numerators obey the same symmetries as their associated color factors.
Such numerators can be used to construct gravity amplitudes using the double-copy procedure.
\item 
amplitude-encoding: the external state dependence (\eg, helicities) of
the numerators is expressed in terms of color-ordered amplitudes.
Such a representation would also be independent of the number of space-time dimensions.
\item
symmetry:  
all of the numerators $n_i$ for a given diagram topology can 
be expressed via permutations of the arguments of a single function.
This virtue is thus one of economy, and also makes the gravity amplitudes 
constructed from these numerators manifestly invariant under permutations
of the external legs.
\end{enumerate}
Kinematic numerators that satisfy all three features are dubbed ``virtuous.''
\para

As we saw in the previous section, numerators that satisfy the
first virtue are expressed in terms
of an independent basis of half-ladder numerators 
${\bf n}_{1 \gamma n } (\range)$ via \eqn{nMn}.
Color-ordered amplitudes are expressed as in terms of these as
\be
\AYM_\alpha (\range) 
~=~  \sum_{\gamma \in S_{n-2}}  
m( \alpha| 1 \gamma n) 
~{\bf n}_{1 \gamma n } (\range) \,.
\label{Amnrepeat}
\ee
Were we able to invert this equation, we would possess numerators
that also satisfy the second virtue of amplitude-encoding.
Because $\det m= 0$, \eqn{Amnrepeat} has no unique inverse;  
there exists rather a family of generalized inverses \cite{BenIsrael,Boels:2012sy},
each corresponding to a particular gauge choice for ${\bf n}_{1 \gamma n }$.
One such choice is given by \cite{Cachazo:2013iea}
\ba
{\bf n}_{1\gamma(2)\cdots\gamma(n-1)n}(\range) &=& 0, \qquad \qquad \gamma(n-1)\neq n-1
\nl
{\bf n}_{1\gamma(2)\cdots\gamma(n-2)n-1,n} (\range)  &=&
-~\sum_{\beta\in S_{n-3}}S[\gamma|\beta] 
\AYM (1,\beta(2), \cdots, \beta(n-2) ,n,n-1)\,.
\label{gaugechoice}
\ea
This choice, however, does not satisfy the third virtue of symmetry.
\para

Next, we describe how one can generate a symmetric representation of 
numerators starting from a nonsymmetric representation,
such as that given in \eqn{gaugechoice}.\footnote{We thank Freddy Cachazo for 
suggesting this approach.}
We show in the appendix that, 
given one set of numerators
${\bf n}_{\alpha} (\range)$,
one can use an arbitrary permutation $\beta$ acting on the 
external legs to generate another set of valid numerators 
\be
{\bf n}'_{\alpha} (\range)
~=~ {\bf n}_{\beta^{-1} \alpha} (\beta(1), \cdots, \beta(n) ) 
\ee
\ie, these also satisfy \eqn{Amnrepeat}.
If  we average over all permutations $\beta \in S_n$,
the resulting function
\be
{\bf n} (\range) 
~\equiv~
 {\bf n}^{\rm sym} (\range) 
~=~ {1 \over n!} \sum_{\beta \in S_n}  {\bf n}_{\beta^{-1} } (\beta(1), \cdots, \beta(n) ) 
\label{symmetrized}
\ee
will be symmetric; \ie, all half-ladder numerators are given by permutations  
of the arguments of the single function ${\bf n} (\range) $.
Symmetric numerator functions for topologies other than the half-ladder are then
obtained from ${\bf n} (\range) $ via \eqn{nMn}.
Since the individual sets of numerators
${\bf n}'_{\alpha} (\range)$ satisfy \eqn{Amnrepeat},
so does the average (\ref{symmetrized}):
\be
\AYM (\alpha(1), \cdots, \alpha(n)) 
 ~=~   
\sum_{\gamma \in S_{n-2}}  
~m( \alpha| 1 \gamma n) 
~{\bf n} (1, \gamma(2), \cdots, \gamma(n-1), n )\,.
\label{validvirtue}
\ee
Hence we have constructed a representation of kinematic numerators satisfying all three virtues of Broedel and Carrasco.
\para

We apply this procedure in the following subsections to obtain 
explicit expressions for virtuous numerators for four- and five-point amplitudes.
The expressions become increasingly lengthy for higher-point amplitudes, but
the important point is that there exists a constructive proof of the existence of 
virtuous numerators for all tree-level $n$-gluon amplitudes.
Whether such numerators are unique remains an open question.
\para

In a recent paper \cite{Fu:2014pya}, Fu, Du, and Feng
also gave a prescription for obtaining virtuous numerators
based on a similar symmetrization strategy applied to 
a KLT expression for the gauge-theory amplitude. 
It seems likely that this gives the same results as \eqn{symmetrized}.
We will compare specific results below.

\subsection{Four-point symmetric kinematic numerators}

For the four-point amplitude, there is only one topology
for color factors, the half ladder. 
A symmetric half-ladder numerator ${\bf n}(1,2,3,4)$  must satisfy \eqn{validvirtue}
which in this case becomes
\be
A(1,2,3,4) ~=~   \left( {1 \over s_{12}} + {1 \over s_{14}} \right) {\bf n}(1,2,3,4)     
		~+~ {1 \over s_{14}} {\bf n}(1,3,4,2)   \,.
\label{fourpointcolorordered}
\ee
\Eqn{gaugechoice} yields
a nonsymmetric amplitude-encoded BCJ representation 
for independent half-ladder numerators
\ba
{\bf n}_{1234}(i,j,k,l) &=& - s_{ij} \AYM (i,j,l,k)\,,  \nl
{\bf n}_{1324}(i,j,k,l) &=& 0 
\label{fourpointnonsym}
\ea
with the remaining numerators given by \eqn{nMn}, \ie, by Jacobi identities and numerator symmetries.
We obtain a virtuous representation by summing over all permutations (\ref{symmetrized})
\be
{\bf n}(1,2,3,4) ~=~
{1 \over 12} {(s_{12} + s_{34})} \left[  
A(1,2,3,4) -A(1,3,4,2) 
\right]
+{1 \over 12} {(s_{13} +s_{24} -s_{14} - s_{23})}  
A(1,4,2,3)  \,.
\label{fourpointvirtue}
\ee
We have reduced the number of terms by using
the cyclic invariance of $A(1,2,3,4)$
as well as the reversal symmetry $A(1,2,3,4)= A(4,3,2,1)$.
This expression manifestly obeys the Jacobi identities
\be
{\bf n}(1,2,3,4) + {\bf n}(1,3,4,2) + {\bf n}(1,4,2,3)  ~=~ 0
\ee
as well as the dihedral symmetries of the half-ladder diagram
\be
{\bf n}(1,2,3,4) ~=~  - {\bf n}(2,1,3,4) ~=~  - {\bf n}(1,2,4,3) ~=~  {\bf n}(4,3,2,1)  \,.
\ee
\Eqn{fourpointvirtue} also satisfies \eqn{fourpointcolorordered}
provided that  the color-ordered amplitudes satisfy the four-point BCJ relations 
\be
s_{ij} A(i,j,k,l) ~=~  s_{ik} A(i,l,j,k) \,.
\label{fourpointBCJ}
\ee
Our expression (\ref{fourpointvirtue}) agrees with
that recently obtained in ref.~\cite{Fu:2014pya}.
It is also in agreement with the shorter expression in ref.~\cite{Broedel:2011pd}
\be
{\bf n}(1,2,3,4) ~=~   {1 \over 3} \left[ s_{12}  A (1, 2, 3, 4)  -  s_{14}  A(1, 4, 2, 3) \right]
\label{fourpointBC}
\ee
once momentum conservation and the Kleiss-Kuijf (subcyclic) identity
$ A(1,2,3,4) + A(1,3,4,2) + A(1, 4, 2, 3) =  0 $
are imposed.
In fact, the three virtues listed above are sufficient to uniquely determine
the four-point numerator.
\para

The BCJ relations (\ref{fourpointBCJ}) can be used to rewrite \eqn{fourpointBC} as 
\be
{\bf n}(1,2,3,4) ~=~   {1 \over 3} s_{12} \left[ A(1,2,3,4) - A(1, 3, 4, 2) \right]  \,.
\ee
We could further use the BCJ relations to write 
${\bf n}(1,2,3,4)$
in terms of a single color-ordered amplitude,
but only at the price of having kinematic invariants in the denominator.

\subsection{Five-point symmetric numerators}

For the five-point amplitude, there is again only one topology
for the color factors, the half ladder. 
A symmetric half-ladder numerator ${\bf n}(1,2,3,4,5)$  satisfies \eqn{validvirtue}, 
which after using Jacobi identities is equivalent to 
\be
A(1,2,3,4,5) ~=~  
 {{\bf n}(1,2,3,4,5)    \over s_{12}s_{45}}  
~+~{{\bf n}(2,3,4,5,1)    \over s_{23}s_{51}}  
~+~{{\bf n}(3,4,5,1,2)    \over s_{34}s_{12}}  
~+~{{\bf n}(4,5,1,2,3)    \over s_{45}s_{23}}  
~+~{{\bf n}(5,1,2,3,4)    \over s_{51}s_{34}}   \,.
\label{fivepointcolorordered}
\ee
\Eqn{gaugechoice} yields
a nonsymmetric amplitude-encoded BCJ representation 
for independent half-ladder numerators
\ba
{\bf n}_{12345}(i,j,k,l,m) &=&  - s_{ij} (s_{ik}+s_{jk} )  A(i,j,k,m,l) - s_{ij} s_{ik} A (i,k,j,m,l)\,, \nl
{\bf n}_{13245}(i,j,k,l,m) &=&  - s_{ij} s_{ik}           A(i,j,k,m,l)  - s_{ik} (s_{ij}+s_{jk} )   A(i,k,j,m,l)\,,  \\
{\bf n}_{14235}(i,j,k,l,m) &=&   0\,, \nl
{\bf n}_{12435}(i,j,k,l,m) &=& 0\,, \nl
{\bf n}_{14235}(i,j,k,l,m) &=& 0\,, \nl
{\bf n}_{14325}(i,j,k,l,m) &=&0
\nonumber
\ea
where the other half-ladder numerators are obtained using \eqn{nMn}.
Again, we obtain a virtuous representation by summing over all permutations (\ref{symmetrized})
\ba
{\bf n}(1,2,3,4,5) &=&
{1 \over 60} \Big[ 
\left(s_{1 2} s_{1 3} + 2 s_{1 2} s_{2 3} + 2 s_{3 4} s_{4 5} + s_{3 5} s_{4 5}                                                          
\right) A(1, 2, 3, 4, 5) \nl
&+&\left(s_{1 3} s_{1 4} + s_{1 2} s_{1 5} + s_{2 3} s_{2 4} + 2 s_{2 3} s_{3 4} + s_{2 4} s_{3 4} + s_{2 5} s_{3 5} + s_{1 5} s_{4 5}      
\right) A(1, 4, 3, 2, 5) \nl
&+&\left(s_{1 3} s_{1 4} + s_{2 3} s_{2 4} + s_{1 2} s_{2 5} + 2 s_{1 3} s_{3 4} + s_{1 4} s_{3 4} + s_{1 5} s_{3 5} + s_{2 5} s_{4 5}      
\right) A(1, 3, 4, 2, 5) \nl
&+&\left(s_{1 2} s_{1 4} - 2 s_{1 2} s_{1 5} + 2 s_{1 2} s_{2 4} - s_{1 2} s_{2 5} - s_{3 4} s_{4 5} + s_{3 5} s_{4 5}                      
\right) A(1, 2, 4, 3, 5) \nl
&+&\left(s_{1 2} s_{1 4} + s_{1 3} s_{1 5} + s_{2 3} s_{2 5} + s_{2 4} s_{3 4} + 2 s_{2 3} s_{3 5} + s_{2 5} s_{3 5} + s_{1 4} s_{4 5}      
\right) A(1, 4, 2, 3, 5) \nl
&+&\left(s_{1 2} s_{1 3} - s_{1 2} s_{2 3} - s_{1 4} s_{4 5} - 2 s_{1 5} s_{4 5} + 2 s_{2 4} s_{4 5} + s_{2 5} s_{4 5}                      
\right) A(1, 3, 2, 4, 5) \nl
&+&\left(2 s_{1 2} s_{1 3} + s_{1 2} s_{2 3} + 2 s_{3 4} s_{4 5} + s_{3 5} s_{4 5}                                                          
\right) A(1, 2, 5, 4, 3) \nl
&+&\left(s_{1 2} s_{1 3} - s_{1 2} s_{2 3} - 2 s_{1 4} s_{4 5} - s_{1 5} s_{4 5} + s_{2 4} s_{4 5} + 2 s_{2 5} s_{4 5}                      
\right) A(1, 4, 5, 2, 3) \nl
&+&\left(-s_{1 3} s_{1 5} - s_{1 2} s_{2 4} - s_{2 3} s_{2 5} - s_{1 4} s_{3 4} - 2 s_{1 3} s_{3 5} - s_{1 5} s_{3 5} - s_{2 4} s_{4 5}   
\right) A(1, 3, 5, 2, 4) \nl
&+&\left(2 s_{1 2} s_{1 4} - s_{1 2} s_{1 5} + s_{1 2} s_{2 4} - 2 s_{1 2} s_{2 5} - s_{3 4} s_{4 5} + s_{3 5} s_{4 5}                      
\right) A(1, 2, 5, 3, 4) \nl
&+&\left(s_{1 2} s_{1 3} + 2 s_{1 2} s_{2 3} + s_{3 4} s_{4 5} + 2 s_{3 5} s_{4 5}                                                          
\right) A(1, 4, 5, 3, 2) \nl
&+&\left(2 s_{1 2} s_{1 3} + s_{1 2} s_{2 3} + s_{3 4} s_{4 5} + 2 s_{3 5} s_{4 5}                                                          
\right) A(1, 3, 5, 4, 2)
\Big]
\label{fivepointvirtue}
\ea
where we have used the reversal property 
$ A(1,2,3,4,5)= -A(5,4,3,2,1) $
together with cyclic invariance of $A(1,2,3,4,5)$.
This expression automatically obeys the dihedral symmetry of the half-ladder diagram
\be
{\bf n}(1,2,3,4,5) ~=~  - {\bf n}(2,1,3,4,5) ~=~  - {\bf n}(1,2,3,5,4) ~=~ - {\bf n}(5,4,3,2,1) 
\label{fivepointdihedral}
\ee
as well as all the Jacobi identities.
\para

\Eqn{fivepointvirtue} agrees with the shorter expression 
given in Broedel and Carrasco \cite{Broedel:2011pd}
after imposing momentum conservation and BCJ relations
on the color-ordered amplitudes.
The Broedel-Carrasco expression is virtuous, 
but to show that it obeys \eqn{fivepointdihedral}
requires the imposition of momentum conservation and BCJ relations,
whereas for \eqn{fivepointvirtue} the dihedral symmetry is manifest.
Both the Broedel-Carrasco expression and \eqn{fivepointvirtue} 
satisfy \eqn{fivepointcolorordered} only after momentum conservation and the BCJ relations
are imposed.
\para

We can re-express \eqn{fivepointvirtue} in the Kleiss-Kuijf basis
to find
\ba
{\bf n}(1,2,3,4,5) &=&
{1 \over 10} \Big[ 
 s_{12} \left(s_{13}+s_{23}\right)+s_{45}\left(s_{34}+s_{35}\right) 
\Big] A(1, 2, 3, 4, 5)
\nl &+&
{1 \over 60} \Big[ 
s_{12} \left(s_{13}+2 s_{14}-s_{23}-2 s_{25}\right)
+s_{13} \left(s_{14}-s_{15}-2 s_{35}\right)
+s_{23} \left(s_{24}-s_{25}+2 s_{34}\right)
\nl &&
+s_{34} \left( s_{24}-s_{14} \right)
+s_{35} \left(s_{25}-s_{15}\right) 
+s_{45} \left( -2 s_{14}+2 s_{25}-s_{34}+s_{35} \right)
\Big]
A(1, 4, 3, 2, 5)
\nl &+&
{1 \over 60} \Big[ 
s_{12} \left(3 s_{13}-s_{24}+s_{25}\right)+s_{13} \left(s_{14}-s_{15}+2 s_{34}-2 s_{35}\right)
+ s_{23} \left( s_{24}-s_{25} \right)
\nl &&
+ s_{45} \left(-2 s_{14}-s_{15}+3 s_{25}+2 s_{34}+s_{35}\right) 
\Big]
A(1, 3, 4, 2, 5)
\nl &+&
{1 \over 20} \Big[ 
s_{12} \left(s_{13}+s_{14}-s_{15}+s_{23}+s_{24}-s_{25}\right)+2 s_{35} s_{45}
\Big]
A(1, 2, 4, 3, 5)
\nl &+&
{1 \over 60} \Big[ 
 s_{12} \left(s_{13}+3 s_{14}-s_{15}+2 s_{23}-2 s_{25}\right)
+s_{14} \left(s_{45}-s_{34}\right)
\nl &&
+ s_{24} \left(s_{34}-s_{45}\right)+s_{35} \left(-2 s_{13}-s_{15}+2 s_{23}+s_{25}+3 s_{45}\right)
\Big]
A(1, 4, 2, 3, 5)
\nl &+&
{1 \over 20} \Big[ 
 2 s_{12} s_{13}+
s_{45}\left(-s_{14}-s_{15}+s_{24}+s_{25}+s_{34}+s_{35}\right) 
\Big]
A(1, 3, 2, 4, 5) \,.
\label{kkbasis}
\ea
This expression differs from the result given in ref.~\cite{Fu:2014pya},
but that result apparently contains some typographical errors which will be corrected
in a revised version\footnote{Private communication}.
We emphasize that \eqn{kkbasis} manifestly obeys Jacobi identities and dihedral symmetries
using only the Kleiss-Kuijf relations and the cyclic and reversal properties of the color-ordered
amplitudes (but not momentum conservation or BCJ relations).  
\para

Fu et al. \cite{Fu:2014pya} also give explicit, rather lengthy, expressions 
for virtuous numerators for six-point amplitudes.

\section{Virtuous dual-trace functions}
\setcounter{equation}{0}

The color factors $c_i$ and the kinematic numerators $n_i(\range)$ 
play dual roles in the gauge-theory amplitude (\ref{gluonamp}). 
Bern and Dennen \cite{Bern:2011ia} proposed that a role dual
to the traces of generators $\Tr[\alpha]$ in \eqn{tracedecomp} could be played
by a function $\tau_\alpha(\range)$ of the variables $p_j$ and $\varepsilon_j$
that is related to $n_i(\range)$
in the same way (\ref{colortrace}) that $\Tr[\alpha]$ is related to $c_i$: 
\be
n_i (\range)  ~=~
\sum_{\alpha  \in \SmodZ}
M_{i \alpha} \, \tau_\alpha (\range)   \,.
\label{nMtau}
\ee
Just as the Jacobi identities for the color factors (\ref{cjacobi}) 
can be expressed as $\sum_i \ell_i c_i  = 0$,
where $\ell_i$ are left null vectors of the rank $(n-2)!$ matrix $M_{i\alpha}$,
so the Jacobi identities for the kinematic numerators (\ref{njacobi}) 
are expressed as $\sum_i \ell_i n_i  = 0$,
and will therefore be automatically satisfied by \eqn{nMtau}.
Using \eqn{nMtau}, together with \eqns{colorordered}{doublepartial}, the color-ordered amplitudes can be written
in terms of the dual-trace functions as 
\be
\AYM_\alpha (\range) ~=~
\sum_{\beta \in \SmodZ}  
m(\alpha|\beta) 
~ \tau_{\beta} (\range) \,.
\label{Amtau}
\ee
If we could invert this equation, we would have
a prescription for an amplitude-encoded dual-trace function, 
i.e. in which the dependence on the external states is expressed
through the color-ordered amplitudes. 
Since $m(\alpha|\beta)$ has rank $(n-3)!$, 
there is no unique inverse, but rather a family of generalized inverses, 
each corresponding to a particular
gauge choice imposed on the dual-trace functions.
\para

First we consider the symmetries of the dual-trace functions $\tau_\alpha$.
Because $M_{i\alpha} = M_{i\alpha'}$ 
when $\alpha$ and $\alpha'$ are related by cyclic permutations,
and $ M_{i\alpha} = (-1)^n M_{i\alpha'} $ 
when $\alpha'$ is the reverse of $\alpha'$,
we can impose these properties on $\tau_\alpha$ without loss of generality
\ba
\tau_{\alpha(1)\alpha(2)\cdots \alpha(n)} (\range)
&= &\tau_{\alpha(2)\alpha(3)\cdots \alpha(1)} (\range) \,,
\label{taucyclic}
\\[2 mm] 
\tau_{\alpha(1) \alpha(2)\cdots \alpha(n)} (\range)
&=& (-1)^n \tau_{\alpha(n) \cdots \alpha(2) \alpha(1)} (\range)
\label{taureverse}
\ea
leaving $(n-1)!/2$ independent dual-trace functions to be determined.
These must satisfy \eqn{nMtau}, which for the independent half-ladder numerators
takes the form
\ba
{\bf n}_{1 \gamma(2) \cdots \gamma(n-1) n} (\range)  &=&  \tau_{ 1[ \gamma(2), [ \cdots, [\gamma(n-1), n] \cdots ]]  } (\range)
\label{ntau}
\\ [2mm]
&=& \tau_{1 \gamma(2) \cdots \gamma(n-1) n}  (\range) + (-1)^n
                 \tau_{1 n \gamma(n-1) \cdots \gamma(2)} (\range)  + \cdots 
\nonumber
\ea
where $+\cdots$ consists of terms 
$\tau_{1\gamma(2)\cdots\gamma(n)}$ 
for which $\gamma(2)\neq n$ and $\gamma(n)\neq n$.
Since there are only  $(n-2)!$ kinematic numerators
${\bf n}_{1 \gamma n} $
and $(n-1)!/2$ dual-trace functions $\tau_\alpha$, 
there remains a great deal of (gauge) freedom in choosing $\tau_\alpha$.
\para

In refs.~\cite{Bern:2011ia,Du:2013sha}, Kleiss-Kuijf relations $\sum_\alpha r_\alpha \tau_\alpha = 0$ 
were imposed on $\tau_\alpha$, reducing the number of independent dual-trace functions to $(n-2)!$.
This is an optional\footnote{
Recall that Kleiss-Kuijf relations follow from the existence of right null 
vectors 
$\sum_\alpha M_{i\alpha}r_\alpha = 0$
of the matrix $M_{i\alpha}$. 
By virtue of \eqn{colorordered}, 
the color-ordered amplitudes $A_\alpha$ satisfy the
relations $\sum_\alpha r_\alpha A_\alpha =0$.
Kleiss-Kuijf relations do not apply to the trace basis $\Tr[\alpha]$ 
and therefore one is not required to impose them on $\tau_\alpha.$ 
For example, see ref.~\cite{Fu:2013qna}.}
gauge choice, however, 
and we choose instead to set to zero all terms of the form 
$\tau_{1\gamma(2)\cdots\gamma(n)}  (\range)$
except 
\ba
\tau_{1\gamma(2)\cdots\gamma(n-2)n-1,n}  (\range)
&=& -{1\over 2}\sum_{\beta\in S_{n-3}}S[\gamma|\beta]A(1,\beta(2), \cdots, \beta(n-2) ,n,n-1),  
\nl
\tau_{1,n,n-1\gamma(n-2)\cdots\gamma(2)}  (\range)
&=& ( -1)^n  \tau_{1\gamma(2)\cdots\gamma(n-2)n-1,n}  (\range) \,.
\label{tauchoice}
\ea
All remaining dual-trace functions follow from cyclic invariance (\ref{taucyclic}).
Note that our gauge choice for $\tau_\alpha$ also implies a particular gauge choice for 
${\bf n}_{1 \gamma n} $, namely, \eqn{gaugechoice}.
According to ref.~\cite{Bern:2011ia}, the ability to express the dual-trace functions in 
terms of kinematic numerators requires us to impose Kleiss-Kuijf relations on $\tau_\alpha$.
If, however, our goal is to write amplitude-encoded dual-trace functions, then this restriction
is not necessary, as we will see explicitly below.
\para

The dual-trace functions defined in \eqn{tauchoice} 
are amplitude-encoded and satisfy \eqn{Amtau},
but they are not symmetric functions in the sense of Broedel and Carrasco.\footnote{In the language of 
ref.~\cite{Du:2013sha}, they do not have a natural relabeling property.}
In the appendix, it is shown that we can
follow the same procedure as in the previous section 
to generate from \eqn{tauchoice} a symmetric dual-trace function 
\be
\tau (\range ) ~\equiv~  
{1 \over n!} \sum_{\beta \in S_n}  \tau_{\beta^{-1} } (\beta(1), \cdots, \beta(n))  \,.
\label{symmetrictau}
\ee
This then provides a constructive definition 
for a virtuous dual-trace function for tree-level $n$-point amplitudes,
proving that such a representation exists for all $n$.
The symmetric dual-trace function can be used to express the $n$-gluon amplitude as
\be
\cAYM (\range)
~=~
\sum_{\alpha \in \SmodZ}  \sum_{\beta \in \SmodZ}  
\Tr[\alpha]
~m(\alpha|\beta) 
~\tau (\beta)
\label{Tmtau}
\ee
where $\tau(\beta) \equiv \tau(\beta(1), \beta(2), \cdots, \beta(n))$.
\para

By substituting \eqn{nMtau} into (\ref{gravamp}),
the scattering amplitude for gravitons may be written 
\be
\cAgv (\range)
~=~
\sum_{\alpha \in \SmodZ}  \sum_{\beta \in \SmodZ}  
\tau (\alpha) ~ m(\alpha|\beta) ~ \tau (\beta) \,.
\ee
This may equivalently be obtained by replacing $\Tr[\alpha]$ with $\tau_\alpha$ in \eqn{Tmtau}.
The gravity amplitude may also be written \cite{Bern:2011ia} 
\be
\cAgv (\range)
~=~
\sum_{\alpha \in \SmodZ}  
\tau (\alpha) ~\AYM (\alpha)
\ee
which has the nice feature of demonstrating that
the gravity amplitude is {\it manifestly} invariant under an arbitrary
permutation $\beta \in S_n$:
\be
\cAgv (\beta )
~=~
\sum_{\alpha \in \SmodZ}  
\tau (\beta \alpha)
~\AYM (\beta \alpha)
~=~
\sum_{\alpha' \in \SmodZ}  
\tau (\alpha')
~\AYM (\alpha')
~=~
\cAgv (\id)\,.
\ee
\para

In the following subsections, we compute the symmetric dual-trace function (\ref{symmetrictau}) 
explicitly for four- and five-point amplitudes. 
Despite being symmetric, our expressions do not satisfy the Kleiss-Kuijf relations,
illustrating that the virtues of Broedel and Carrasco
are not sufficient to single out a unique dual-trace function.
\para

\subsection{Four-point symmetric dual-trace functions}

For the four-gluon amplitude, \eqn{tauchoice} yields a nonsymmetric representation 
for the dual-trace function
\ba
\tau_{1234}(i,j,k,l) &=& - {1 \over 2} s_{ij} \AYM (i,j,l,k)   \,,
\nl
\tau_{1342}(i,j,k,l) &=& 0 \,,
\nl
\tau_{1423}(i,j,k,l) &=& 0
\label{fournonsymtau}
\ea
with the remaining $\tau_\alpha$ given by \eqn{taucyclic}.
We now use \eqn{symmetrictau} to generate a symmetric dual-trace function, obtaining
\be
\tau(1,2,3,4) ~=~
- \frac{1}{24}
\Big[  
\left(s_{12}+s_{34}\right) A(1,3,4,2) + \left(s_{14}+s_{23}\right)
A(1,4,2,3) 
\Big]
\label{fourpointtau}
\ee
where we have used 
$A(1,2,3,4)= A(4,3,2,1)$
and the cyclic invariance of $A(1,2,3,4)$.
\Eqn{fourpointtau} manifestly obeys
$\tau(1,2,3,4)= \tau(2,3,4,1)$ and $\tau(1,2,3,4)= \tau(4,3,2,1)$;
however, 
$\tau(1,2,3,4) + \tau(1,3,4,2) + \tau(1,4,2,3)$ does not vanish;
\ie, 
\eqn{fourpointtau} does not satisfy the Kleiss-Kuijf relations.
When substituted into \eqn{ntau}, which takes the form 
\ba
{\bf n}(1,2,3,4) &=&  2 \left[ \tau(1,2,3,4)  -   \tau(1,2,4,3)  \right] \,,
\label{fourntau}
\ea
\eqn{fourpointtau} yields precisely the symmetric kinematic numerator found in \eqn{fourpointvirtue}.
\Eqn {fourpointtau} can be written more briefly as 
\be
\tau (1,2,3,4)
~=~  - {1 \over 6} s_{12}  A(1, 3, 4,2) 
\ee
by using momentum conservation and the BCJ relations (\ref{fourpointBCJ}).
\para

Bern and Dennen proposed that the four-point dual-trace function takes the form \cite{Bern:2011ia}
\be
\tau_{BD} (1,2,3,4) ~=~
{1 \over 6} \left[ {\bf n}(1,2,3,4) + {\bf n}(2,3,4,1)  \right] \,.
\label{BernDennentau}
\ee
The expression for $\tau$ given in ref.~\cite{Du:2013sha} is equivalent to \eqn{BernDennentau}.
\Eqn{BernDennentau}
manifestly obeys
$\tau(1,2,3,4)= \tau(2,3,4,1)$ and $\tau(1,2,3,4)= \tau(4,3,2,1)$,
and in addition satisfies the Kleiss-Kuijf relation.
To compare our expression with that of Bern and Dennen, 
we substitute the symmetric numerator (\ref{fourpointvirtue}) into \eqn{BernDennentau} to find
\ba 
&&\tau_{BD} (1,2,3,4) ~-~ \tau(1,2,3,4)
\\
&&~~~~ = ~\frac{1}{36}
\Big[  
\left(s_{12}+s_{14}\right) A(1,2,3,4) + 
\left(s_{12}+s_{13}\right) A(1,3,4,2) + 
\left(s_{13}+s_{14}\right) A(1,4,2,3) 
\Big] \,.
\nonumber
\ea
The difference is ``pure gauge'': it vanishes when substituted into \eqn{fourntau},
and therefore does not contribute  to ${\bf n}(1,2,3,4)$.
Hence we see that imposing the three virtues of Broedel and Carrasco
on the dual-trace function is not sufficient to 
determine it uniquely.
\para

To further elucidate the difference between \eqns{fourpointtau}{BernDennentau}, 
we rewrite \eqn{Amtau} 
as a matrix equation
\be
\begin{pmatrix} 
A_{1234} \\
A_{1342} \\
A_{1423} 
\end{pmatrix}
~=~ m  
\begin{pmatrix} 
\tau_{1234} \\
\tau_{1342} \\
\tau_{1423} 
\end{pmatrix},
\qquad\qquad
m ~=~-~{2\over s t u} \begin{pmatrix}   u \\ t \\ s \end{pmatrix} \begin{pmatrix}   u &t & s\end{pmatrix}
\label{matrix}
\ee
where $s=s_{12}$, $t=s_{14}$, and $u=s_{13}$.
As expected, $m$ has rank $(4-3)! = 1$.
It therefore does not possess a unique inverse, but rather a 
family of generalized inverses $m^+$ 
\be
\begin{pmatrix} 
\tau_{1234} \\
\tau_{1342} \\
\tau_{1423} 
\end{pmatrix}
~=~ m ^+
\begin{pmatrix} 
A_{1234} \\
A_{1342} \\
A_{1423} 
\end{pmatrix},
\qquad\qquad m m^+ m= m \,.
\ee
A generalized inverse must satisfy $m m^+ m= m$  \cite{BenIsrael,Boels:2012sy},
which guarantees that the resulting $\tau_\alpha$ will satisfy \eqn{matrix}.
\para

Different gauge choices for $\tau$ correspond to different generalized inverses.
The nonsymmetric representation (\ref{fournonsymtau})
corresponds to 
\be
m_{\rm nonsym}^+ ~=~-~{1\over 2} \begin{pmatrix}   0 & s & 0 \\ 0 & 0 & 0 \\ 0 & 0 & 0 \end{pmatrix}  \,.
\ee
The symmetric dual-trace function (\ref{fourpointtau}) corresponds to 
\be
m_{\rm sym}^+ ~=~-~{1\over 12} \begin{pmatrix}   0 & s & t \\ s & 0 & u \\ t & u & 0 \end{pmatrix}  
\ee
whereas  the Bern-Dennen dual-trace function corresponds to 
\be
m_{BD}^+ ~=~{1\over 36} 
\begin{pmatrix}
-u & -3 s-t & -3t-s \\
 -3 s-u & -t & -3u-s \\
 -3 t-u & -3u-t u & -s \\
\end{pmatrix} \,.
\ee
All three generalized inverses satisfy $m m^+ m= m$.

\subsection{Five-point symmetric dual-trace functions} 

For the five-gluon amplitude, \eqn{tauchoice} yields a nonsymmetric representation 
\ba
\tau_{12345}(i,j,k,l,m) &=&  
- {1\over 2} \left[ s_{ij} (s_{ik}+s_{jk} )  A(i,j,k,m,l) + s_{ij} s_{ik} A (i,k,j,m,l)\right]\,, \nl
\tau_{13245}(i,j,k,l,m) &=&  
- {1\over 2} \left[ s_{ij} s_{ik}           A(i,j,k,m,l)  + s_{ik} (s_{ij}+s_{jk} )   A(i,k,j,m,l) \right] 
\label{fivepointnonsymtau}
\ea
with all remaining dual-trace functions set to zero, except for those related to \eqn{fivepointnonsymtau} by \eqn{taucyclic}.
We now use \eqn{symmetrictau} to generate a symmetric dual-trace function, obtaining 
\ba
\tau(1,2,3,4,5) &=&
\frac{1}{120} 
\Big[ 
(s_{23} s_{24}+s_{34} s_{24}+2 s_{23} s_{34}) A(1, 4, 3, 2, 5)
+(s_{23} s_{24}+s_{12} s_{25}) A(1, 3, 4, 2, 5)
\nl &-&
(2 s_{12} s_{15}+s_{25} s_{15}+s_{12} s_{25}) A(1, 2, 4, 3, 5)
+(s_{24} s_{34}+s_{14} s_{45}) A(1, 4, 2, 3, 5)
\nl &-&
(s_{14} s_{15}+2 s_{45} s_{15}+s_{14} s_{45}) A(1, 3, 2, 4, 5)
+(s_{34} s_{35}+s_{45} s_{35}+2 s_{34} s_{45}) A(1, 2, 5, 4, 3)
\nl &+&
(s_{12} s_{13}+s_{14} s_{15}) A(1, 4, 5, 2, 3)
+(s_{15} s_{25}+s_{35} s_{45}) A(1, 2, 5, 3, 4)
\nl &+&
(s_{12} s_{13}+s_{23} s_{13}+2 s_{12} s_{23}) A(1, 4, 5, 3, 2)
-(s_{13} s_{23}+s_{34} s_{35}) A(1, 3, 5, 4, 2) 
\Big]
\label{fivepointtau}
\ea
where we have used
$ A(1,2,3,4,5)= -A(5,4,3,2,1) $
together with cyclic invariance of $A(1,2,3,4,5)$.
This expression manifestly satisfies $\tau(1,2,3,4,5)=\tau(2,3,4,5,1)$
and $\tau(1,2,3,4,5) = - \tau(5,4,3,2,1)$, but not the Kleiss-Kuijf relations.
It therefore differs from the expressions given 
in refs. \cite{Bern:2011ia,Du:2013sha}, 
which do satisfy the Kleiss-Kuijf relations.
\para

\Eqn{fivepointtau} yields precisely the symmetric kinematic numerator \eqn{fivepointvirtue}
when substituted into \eqn{fiventau}, which takes the form
\ba
{\bf n}(1,2,3,4,5) &=&
2 \left[\tau(1,2,3,4,5)  + \tau(1,2,5,4,3) + \tau(1,4,5,3,2) + \tau(1,3,5,4,2) \right] \,.
\label{fiventau}
\ea
Finally, the Kleiss-Kuijf relations for the color-ordered amplitudes can be used to rewrite \eqn{fivepointtau} 
as
\ba
{\bf \tau}(1,2,3,4,5) &=&
{1 \over 120} \Big[ 
(
 s_{12} s_{13}+2 s_{12} s_{23}+2 s_{34} s_{45}+s_{35} s_{45}
) A(1, 2, 3, 4, 5)
\nl &+&
(
 s_{12} s_{13}+s_{14} s_{15}+s_{23} s_{24}+s_{15} s_{25}+2 s_{23} s_{34}+s_{24} s_{34}+s_{35} s_{45}
)A(1, 4, 3, 2, 5)
\nl &+&
(
 s_{12} s_{13}+s_{14} s_{15}+s_{23} s_{24}+s_{12} s_{25}+s_{34} s_{35}+2 s_{34} s_{45}+s_{35} s_{45}
)A(1, 3, 4, 2, 5)
\nl &+&
(
 s_{12} s_{13}-2 s_{12} s_{15}+2 s_{12} s_{23}-s_{12} s_{25}-s_{34} s_{35}+s_{35} s_{45}
) A(1, 2, 4, 3, 5)
\nl &+&
(
 s_{12} s_{13}+s_{23} s_{13}+2 s_{12} s_{23}+s_{15} s_{25}+s_{24} s_{34}+s_{14} s_{45}+s_{35} s_{45}
)A(1, 4, 2, 3, 5)
\nl &+&
(
 s_{12} s_{13}-s_{23} s_{13}-s_{14} s_{45}-2 s_{15} s_{45}+2 s_{34} s_{45}+s_{35} s_{45}
)A(1, 3, 2, 4, 5)
\Big]\,.
\nl
\ea

\section{Discussion}
\setcounter{equation}{0}

In this paper, we have offered a constructive procedure for computing
virtuous kinematic numerators for $n$-point gauge-theory scattering amplitudes;
that is, numerators that simultaneously satisfy color-kinematic duality,
are expressed in terms of color-ordered amplitudes, 
and are symmetric in the sense of Broedel and Carrasco.
We have presented explicit expressions
for four- and five-point amplitudes, 
which although somewhat lengthy, 
have the advantage of manifestly obeying the Jacobi identities and 
diagram symmetries without having to invoke the BCJ relations 
for the color-ordered amplitudes.
Our results are equivalent (upon using the BCJ relations)
to other virtuous expressions in the literature,
suggesting the possibility that the three virtues of Broedel and Carrasco
are sufficient to uniquely determine the kinematic numerators,
although we do not have a proof of this.
\para

We have also applied this procedure to 
compute symmetric, amplitude-encoded dual-trace functions $\tau$ for $n$-point amplitudes,
presenting explicit expressions for four- and five-point amplitudes.
In this case, the results are not uniquely determined by these criteria alone.
In particular, symmetric expressions for $\tau$ obtained by other authors 
additionally satisfy (optional) Kleiss-Kuijf relations, whereas our results do not.
While it is possible that a constructive procedure could be found to 
generate virtuous dual-trace functions
for $n$-point amplitudes that also satisfy Kleiss-Kuijf relations,
it is also possible that an alternative criterion
to single out a unique dual-trace function
may be more useful or natural.

\section*{Acknowledgments}
I am grateful to Johannes Broedel for useful correspondence.
I especially wish to thank Freddy Cachazo 
and Ellis Yuan for crucial conversations
at the outset of this project. 
This research was supported in part by the NSF under grant no. PHY10-67961.

\appendix

\section{Generating symmetric kinematic numerators}
\setcounter{equation}{0}

In this appendix, we consider the effect of an arbitrary permutation $\beta$
on the momentum, polarization, and color of the external particles
of the $n$-gluon amplitude:
\be
\cAYM (\beta) ~\equiv~ \cAYM  (\beta(1), \beta(2), \cdots, \beta(n) ) \,.
\ee
Of course, the full bosonic amplitude is invariant under permutations:
$ \cAYM (\beta) = \cAYM(\id)$,
where $\id$ denotes the identity permutation.
We will explore the consequences of this on the color-ordered amplitudes and kinematic
numerators below.
\para

First, we rewrite the trace decomposition (\ref{tracedecomp}) of
the $n$-gluon amplitude  as a sum over all permutations
of the external legs
\be
\cAYM  (\range)
~=~ {1 \over n} \sum_{\alpha \in S_n} \Tr[ \alpha]~ \AYM_{\alpha} (\range)  
\label{tracedecompovern}
\ee
dividing by $n$ to correct for the overcounting.
Since $\Tr[\alpha]=\Tr[\alpha']$ 
when $\alpha$ and $\alpha'$ are related by cyclic permutations,
we impose $ \AYM_{\alpha} (\range)  = \AYM_{\alpha'} (\range) $
on the color-ordered amplitudes without loss of generality. 
The effect of the permutation $\beta$ on the trace decomposition 
(\ref{tracedecompovern}) is 
\be
\cAYM  (\beta) ~=~ {1 \over n} \sum_{\alpha \in S_n} \Tr[\beta\alpha]~ \AYM_\alpha (\beta) 
             ~=~ {1 \over n} \sum_{\alpha' \in S_n}  \Tr[\alpha'] ~\AYM_{\beta^{-1}\alpha'} (\beta) 
\ee
where $\beta\alpha$ denotes the permutation\footnote{
Hence if $\alpha=(12)$ and $\beta=(23)$ then $\beta\alpha= (132)$,
where we use the cycle notation for permutations.
}
obtained by acting first with $\alpha$ and then with $\beta$.
Since the amplitude $\cAYM (\range)$ is invariant
under permutations of the external legs, it follows that
\be
\sum_{\alpha \in S_n}  \Tr[\alpha]~\AYM_{\beta^{-1}\alpha} (\beta) 
~=~  \sum_{\alpha \in S_n} \Tr[\alpha] ~ \AYM_{\alpha} (\id) \,.
\ee
Since the trace basis is independent (modulo cyclic permutations)
we may conclude that the color-ordered amplitudes obey 
\be
\AYM_{\beta^{-1}\alpha} (\beta)  ~=~ \AYM_\alpha(\id)
\qquad \Rightarrow \qquad
\AYM_{\alpha} (\id) ~=~ \AYM_\id(\alpha) \,.
\ee
Denoting
$\AYM_\id (\range)$ as simply $\AYM (\range)$,
we see that the entire set of color-ordered amplitudes $\AYM_\alpha (\range)$
is obtained by permuting the arguments of this one function
\be
\AYM_\alpha (\range) ~=~ \AYM(\alpha(1), \cdots, \alpha(n))
\ee
Thus, color-ordered amplitudes are symmetric 
in the sense of Broedel and Carrasco \cite{Broedel:2011pd}.
\para

Next, we consider the effect of permuting the arguments of the 
amplitude in the color basis.
We begin by rewriting the $n$-gluon amplitude \eqn{gluonamp} as
\be
\cAYM (\range) 
~=~ \sum_T   {1 \over | \GT | }  
  \sum_{\alpha \in S_n}  
{\cT_\alpha  \nT_\alpha (\range) \over \dT (\alpha) }  \,.
\label{gluonampoverG} 
\ee
Here we have divided up the color factors $c_i$ according to the topology $T$
of the diagram, and arbitrarily assign the color labels 
$\textsf{a}_j$ to the legs of the diagram, 
calling the result $c_\id^{(T)}$.
Then $c_\alpha^{(T)}$ denotes the color factor with the same topology 
but with the color labels on the legs permuted by $\alpha \in S_n$.
Also, $\dT(\range)$  is the denominator associated with  $c_\id^{(T)}$
and $\dT(\alpha) = \dT( \alpha(1), \cdots, \alpha(n))$.
\para

Associated with each topology is a symmetry group $\GT \subset S_n$ 
under which the color factor is invariant
(modulo sign)
\be
\cT_{\alpha\rho} ~=~  \pm \cT_\alpha,
\qquad \rho \in G^{(T)}\,.
\label{csymmetry}
\ee
For example, the symmetries of the half-ladder color factor
\be
{\bf c}_{123\cdots n-2,n-1,n} ~=~ 
-{\bf c}_{213\cdots n-2,n-1,n} ~=~ 
-{\bf c}_{123\cdots n-2,n,n-1} ~=~ 
(-1)^n  {\bf c}_{n,n-1,n-2 \cdots 321} 
\ee
generate the order 8 dihedral group.
In \eqn{gluonampoverG},
we have divided by $|\GT|$, the order of $\GT$,
to correct for the overcounting that results from summing over redundant
diagrams.  
The denominator is invariant under $\GT$
\be
\dT (\alpha) ~=~ \dT (\alpha\rho), \qquad \rho \in G^{(T)}
\ee
and without loss of generality, 
the symmetries of $\cT_\alpha$ may also be imposed on the numerators 
\be
\nT_\alpha (\range)  ~=~  \pm \nT_{\alpha\rho} (\range) , \qquad \rho \in G^{(T)} 
\label{nsymmetry}
\ee
since any nonsymmetric piece will drop out of the sum over permutations.
Thus we can restrict the sum to $\SmodG$ 
\be
\cAYM (\range)
~=~ \sum_T   \sum_{\alpha \in \SmodG}  
{\cT_\alpha  \nT_\alpha (\range) \over \dT (\alpha) }  \,.
\ee
\para

Unlike the trace basis, the set of color factors (modulo $\GT$) does
not constitute an independent basis on account of various Jacobi identities
of the form
\be
c^{(T_1)}_{\alpha_1} + c^{(T_2)}_{\alpha_2} + c^{(T_3)}_{\alpha_3} ~=~0
\ee
where $T_1$, $T_2$, and $T_3$  need not all be equal since Jacobi identities 
may mix color factors with different topologies.
Hence the kinematic numerators $\nT_\alpha$ are not unique
and can undergo generalized gauge transformations.
\para

Now we apply a permutation $\beta$ to the external legs of an amplitude 
expressed in the color basis (\eqn{gluonampoverG}) to obtain
\be
\cAYM (\beta)
~=~ 
\sum_T   {1 \over | \GT| }  \sum_{\alpha \in S_n}  
{\cT_{\beta\alpha}  \nT_{\alpha} (\beta) \over \dT (\beta\alpha) } 
~=~ 
\sum_T   {1 \over | \GT| }  \sum_{\alpha' \in S_n}  
{\cT_{\alpha'}  \nT_{\beta^{-1} \alpha'} (\beta) \over \dT (\alpha') } 
~=~ 
\sum_T    \sum_{\alpha' \in \SmodG}  
{\cT_{\alpha'}  \nT_{\beta^{-1} \alpha'} (\beta) \over \dT (\alpha') } 
\ee
where in the last equation, we used
\be
\nT_{\beta^{-1} \alpha \rho} (\beta) 
~=~ \pm \nT_{\beta^{-1} \alpha } (\beta), 
\qquad \rho \in G^{(T)} 
\ee
which follows from \eqn{nsymmetry}.
The invariance of the full bosonic amplitude $\cAYM (\range)$
under permutations of the external legs implies that
\be
\sum_T    \sum_{\alpha \in \SmodG}  
{\cT_{\alpha}  \nT_{\beta^{-1} \alpha} (\beta) \over \dT (\alpha) } 
~=~ \sum_T   \sum_{\alpha \in \SmodG}  
{\cT_\alpha  \nT_\alpha (\id) \over \dT (\alpha) } 
\label{permutednumerators}
\ee
but we may not conclude that 
$ \nT_{\beta^{-1} \alpha} (\beta)$
is equal to 
$ \nT_{\alpha} (\id)$,
because as stated above the color factors (modulo the symmetries $\GT$) 
do not form an independent basis.
In other words, the kinematic numerators need not be symmetric, 
in the sense of Broedel and Carrasco \cite{Broedel:2011pd}.
In particular, the kinematic numerators given in \eqn{gaugechoice} 
are manifestly {\it not} symmetric because some of them vanish, 
which is not possible if they are all related
by permutations to a single function.
\para

On the other hand, the permutation symmetry of the amplitude gives us
a mechanism for generating new representations for kinematic numerators from
existing ones.  
Specifically,
\eqn{permutednumerators}
demonstrates that, given a valid set of kinematic 
numerators $ \nT_{\alpha} (\id)$,
an arbitrary permutation $\beta$ generates another valid set of numerators
$ {\nT}'_{\alpha} (\id) \equiv \nT_{\beta^{-1} \alpha} (\beta)$.
Furthermore, if the original numerators $ \nT_{\alpha} (\id)$ 
are BCJ numerators,  obeying 
\be
n^{(T_1)}_{\alpha_1} (\id) ~+~ n^{(T_2)}_{\alpha_2} (\id) ~+~ n^{(T_3)}_{\alpha_3} (\id) ~=~0
\ee
then it is also true that 
\be
n^{(T_1)}_{\beta^{-1} \alpha_1} (\beta) ~+~ n^{(T_2)}_{\beta^{-1} \alpha_2} (\beta) ~+~ n^{(T_3)}_{\beta^{-1} \alpha_3} (\beta) ~=~0
\ee
which implies that the new set also obeys the Jacobi identities
\be
{n^{(T_1)}_{\alpha_1}}' (\id) ~+~ {n^{(T_2)}_{\alpha_2}}' (\id) ~+~ {n^{(T_3)}_{\alpha_3}}' (\id) ~=~0
\ee
and hence are BCJ numerators.
\para

Moreover, normalized linear combinations of valid numerators
are also valid numerators.
This suggests that, given a set of nonsymmetric kinematic numerators $n^{(T)}_\alpha (\id)$,
one may generate a set of symmetric numerators
$n^{\sym(T)}_\alpha (\id)$
by simply averaging  over all permutations
\be
n^{\sym(T)}_\alpha (\id)
~\equiv~  {1 \over n!} \sum_{\beta \in S_n}  \nT_{\beta^{-1} \alpha} (\beta)  \,.
\ee
To show that these numerators are indeed symmetric, observe that 
\be
n^{\sym(T)}_\alpha (\gamma) 
~=~ {1 \over n!} \sum_{\beta \in S_n}  \nT_{\beta^{-1} \alpha} (\gamma \beta) 
~=~ {1 \over n!} \sum_{\beta' \in S_n}  \nT_{\beta'^{-1} \gamma \alpha} (\beta') 
~=~ n^{\sym(T)}_{\gamma \alpha} (\id)
\ee
in other words
\be
n^{\sym(T)}_\alpha (\id) ~=~ n^{\sym(T)}_\id (\alpha) \,.
\ee
Thus, all of the numerators of a given topology may be obtained by permuting the arguments 
of one of them. 
To reflect this, we drop the subscript (and the label $sym$) 
and simply define 
\be
\nT (\id ) ~\equiv~  
n^{\sym(T)}_\id (\id) ~=~ 
{1 \over n!} \sum_{\beta \in S_n}  \nT_{\beta^{-1} } (\beta)  \,.
\label{virtuousnum}
\ee
Thus the full amplitude can be written
\be
\cAYM (\range)
~=~ \sum_T   \sum_{\alpha \in \SmodG}  
{\cT_\alpha  \nT (\alpha ) \over \dT (\alpha) } 
\label{virtuousamp}
\ee
in terms of a single symmetric function $\nT (\range)$
for each topology.
Furthermore, using \eqn{nMn},
each $\nT (\range)$ can be expressed in terms of a single symmetric 
function for the half-ladder diagram ${\bf n}(\range)$.
\para

In a similar way, a symmetric dual-trace function can be generated from a nonsymmetric
representation $\tau_\alpha(\id)$.   
The gauge-theory amplitude may be written 
\be
\cAYM (\id) 
~=~ 
{1 \over n^2} 
\sum_{\gamma \in S_n} 
\sum_{\delta \in S_n} 
\Tr[\gamma]   
~m(\gamma|\delta) ~\tau_\delta (\id) 
~=~ 
{1 \over n^2} 
\sum_T   {1 \over | \GT | }  
\sum_{\alpha \in S_n}  
\sum_{\gamma \in S_n} 
\sum_{\delta \in S_n} 
\Tr[\gamma]   
{\MT_{\alpha, \gamma}  \MT_{\alpha, \delta} \over \dT (\alpha) } \tau_\delta (\id) \,.
\ee
Applying a permutation $\beta$ to the external legs, this becomes
\ba
\cAYM (\beta) 
&=&
{1 \over n^2} 
\sum_T   {1 \over | \GT | }  
\sum_{\alpha \in S_n}  
\sum_{\gamma \in S_n} 
\sum_{\delta \in S_n} 
\Tr[\beta\gamma]   
{\MT_{\alpha, \gamma}  \MT_{\alpha, \delta} \over \dT (\beta\alpha) } \tau_\delta (\beta) \,.
\nl
&=&
{1 \over n^2} 
\sum_T   {1 \over | \GT | }  
\sum_{\alpha' \in S_n}  
\sum_{\gamma' \in S_n} 
\sum_{\delta' \in S_n} 
\Tr[\gamma']   
{\MT_{\alpha', \gamma'}  \MT_{\alpha',\delta'} \over \dT (\alpha') } \tau_{\beta^{-1} \delta'} (\beta) 
\nl
&=&
{1 \over n^2} 
\sum_{\gamma \in S_n} 
\sum_{\delta \in S_n} 
\Tr[\gamma]   
~m(\gamma | \delta) ~\tau_{\beta^{-1} \delta} (\beta) 
\ea
where in the second line we used $\MT_{\alpha,\gamma} = \MT_{\beta\alpha,\beta\gamma}$.
Thus $\tau_\delta(\id)$ and $\tau'_\delta (\id) \equiv \tau_{\beta^{-1}\delta} (\beta)$ 
yield the same gauge-theory amplitude.
We may generate a symmetric dual-trace function $\tau(\range)$ by averaging over all permutations $\beta \in S_n$:
\be 
\tau(\delta(1), \cdots, \delta(n))  ~=~ \tau^{\sym}_\delta (\id)
~\equiv~  {1 \over n!} \sum_{\beta \in S_n}  \tau_{\beta^{-1} \delta} (\beta(1), \cdots, \beta(n))  \,.
\ee

\end{document}